# Reservoir Computing Based on Mutually Injected Phase Modulated Semiconductor Lasers as a Monolithic Integrated Hardware Accelerator

Kostas Sozos, Charis Mesaritakis, and Adonis Bogris

*Abstract*—In this paper we propose and numerically study a neuromorphic computing scheme that applies delay-based reservoir computing in a laser system consisting of two mutually coupled phase modulated lasers. The scheme can be monolithic integrated in a straightforward manner and alleviates the need for external optical injection, as the data can be directly applied on the on-chip phase modulator placed between the two lasers. The scheme also offers the benefit of increasing the nodes compared to a reservoir computing system using either one laser under feedback or laser under feedback and optical injection. Numerical simulations assess the performance of the integrated reservoir computing system in dispersion compensation tasks in short-reach optical communication systems. We numerically demonstrate that the proposed platform can recover severely distorted 25 Gbaud PAM-4 signals for transmission distances exceeding 50 km and outperform other competing delay-based reservoir computing systems relying on optical feedback. The proposed scheme, thanks to its compactness and simplicity, can play the role of a monolithic integrated hardware accelerator in a wide range of application requiring high-speed real-time processing.

## I. Introduction

NOWADAYS, machine learning (ML) and artificial intelligence (AI) are two continuously evolving areas that affect all aspects of everyday life, shaping the future of information and communication technologies (ICT) and their penetration in Industry 4.0. Among them, Artificial Neural Networks (ANNs) and their variants like Recurrent Neural Networks (RNNs) have been proved powerful, conquering almost every domain of research and application related to signal processing and big data analytics. Whilst traditional von-Neumann computers are the underlying tools executing the aforementioned algorithms, they tend to be power hungry when trying to solve highly complex computational tasks such as speech or face recognition. T heir use in next generation Internet of Everything architectures as edge computing devices will massively increase the carbon footprint of ICT technologies, especially when distributed machine learning is the target. Thus, new routes to efficient and powerful computing must be considered, and one of them is neuromorphic computing [1].

Photonic platforms are promising candidates to play the role of neuromorphic computing processors as they can provide linear and nonlinear transformations at high speed and low energy consumption [2]. One of the driving forces for powerful photonic processing is reservoir computing (RC). The RC framework is a sub-category of RNNs, consisting of three layers: the input layer, the reservoir and the readout layer. While in typical RNNs all weights are trainable, RC simplifies the training procedure, which takes place in a linear manner only in the readout layer, leaving the reservoir weights random and untrained [3]. Up to date, there have been many demonstrations of photonic RC classified in two main categories: spatially distributed systems [4] and delay-based ones [5]. The delay-based approach introduced in [5], minimizes the hardware required for RC to a single non-linear node via time multiplexing. Many efforts have been made towards speed improvement, power efficiency and miniaturization of such systems, exploiting optical injection and feedback in one or multiple lasers [6], [7], polarization dynamics of VCSELs [8], phase dynamics [9], mutually coupled lasers [10] and multiple longitudinal modes [11], [12]. The introduction of the signal to be processed in all aforementioned systems is accomplished with either optical injection or with the use of current modulation of the laser taking place in the nonlinear dynamical system. The former approach is beneficial provided that the signal to be processed is already in an optical format. However, in most RC demonstrations, especially those related to optical communication applications (see fig. 1), the signal is first detected and transformed to an electrical format and its conversion to an optical signal is costly as it requires the use of an external laser transmitter usually accompanied by an optical modulator, a polarization controller and an optical isolator which makes the system impractical and bulky. Electrical injection through current is an option [13], however optical injection and especially phase modulation can extend the operation bandwidth of the laser under injection either in reservoir computing [9] or in other conventional optical

Manuscript received April 28, 2021; revised July 8, 2021; accepted August 3, 2021. Date of publication August 17, 2021; date of current version August 27, 2021. This work was supported by Hellenic Foundation for Research and Innovation (HFRI) under Grant 2247. *(Corresponding author: Adonis Bogris.)*

Kostas Sozos and Adonis Bogris are with the Department of Informatics and Computer Engineering, University of West Attica, 12243 Egaleo, Greece (e-mail: ksozos@uniwa.gr; abogris@uniwa.gr).

Charis Mesaritakis is with the Department of Information and Communication Systems Engineering, University of the Aegean, 83200 Karlovasi, Samos, Greece (e-mail: cmesar@aegean.gr).

Color versions of one or more figures in this article are available at https://doi.org/10.1109/JQE.2021.3104855.

Digital Object Identifier 10.1109/JQE.2021.3104855

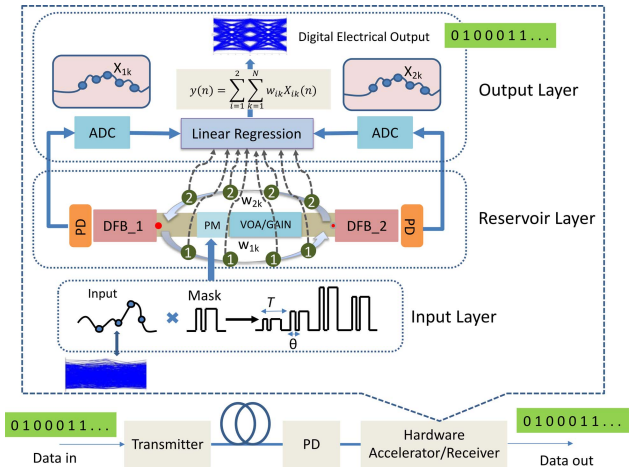

Fig. 1. Fully integrated neuromorphic processor based on mutually coupled Lasers. PM: Phase modulator, VOA: Variable Optical Attenuator, ADC: Analog to Digital Converter, PD: Photodiode, DFB: Distributed Feedback.

signal processing applications [14], which translates to higher speed processing. High speed processing and compactness are the two sides of the same coin in delay-based RC systems, as speed penalty scales up when the optical delay increases. On the other hand, the reduction of overall delay reduces the number of nodes and thus affects processing efficiency. In this context, it is worthwhile to propose a compact RC processor that utilizes electrical injection and at the same time can operate at high processing speeds and at high computational efficiency.

In this paper, we propose a holistic solution employing a scheme comprising two mutually coupled lasers with a phase modulator and a variable optical attenuator (VOA) in the middle, all integrated on the same chip. On the same platform, one can integrate the photodetectors. The optoelectronic system is supported by two ADCs and a light Digital Signal Processing unit that undertakes linear regression. Although an RC system based on mutually coupled lasers has been studied in [10], the specific scheme is less practical as it requires another laser for optical injecting the two mutually coupled sources, two modulators, two masking signals, two polarization controllers and optical isolator. Extensive study of delay coupled systems has been also carried out in [15]; however not for typical diode lasers and always considering different masking for each nonlinear node. On the contrary, the scheme proposed in the present paper does not require optical injection for signal superposition and the complexity it brings, it employs only one masked electrical signal that is superimposed on the phase modulator, it supports high bandwidths as the process relies on optical injection between the two lasers and phase modulation which has proved its efficacy in extending the bandwidth of RC systems [9]. Also, it can operate with half of the delay needed by a typical RC based on an injection locked laser under feedback [16] as it encompasses two spatial RC nodes which translate to doubling the time multiplexed nodes [10]. With this configuration, we launch a black-box RC processor as a hardware accelerator that is very stable and compact and equipped with an electrical input for applying the analog data (phase modulator) and an electrical output which provides the final digital outcome depending on the problem to be solved.

Phase modulation inside the cavity has been also proposed in [17], however for a laser under optical feedback and for much longer optical cavities and smaller modulation rates. Here, for the first time, we evaluate the performance of a RC system comprising coupled lasers without optical external injection and utilizing phase modulation by mitigating the transmission impairments imprinted on a 25 Gbaud PAM-4 signal that has propagated in a single mode fiber. We compare the proposed system with a) the conventional delay-based RC consisting of an injection locked laser subjected to optical feedback [16], b) with a single laser subjected to optical feedback with phase modulation inside cavity [17] and c) with typical linear equalizers [12], [16].

## II. CONCEPTUAL SCHEME AND SYSTEM MODELING

The delay-based RC concept involves, in its simplest form one physical non-linear node whose dynamics are influenced by its own output at a time $t$ in the past. Within one delay interval of length $T$, $N$ equidistant points separated in time by $\theta = T/N$, serve as the virtual nodes [5]. In our scheme, two physical nodes, namely the two lasers separated by distance $T$, are employed. The virtual nodes are distributed in the temporal domain. The system depicted in fig.1 involves two DFB mutually coupled lasers that are phase modulated (PM) with the properly masked PAM-4 signal that has experienced multi km of transmission in a single mode fiber. The masking process refers to the multiplication each symbol of the original signal with a random set of numbers, periodically repeated for every symbol. In this way, the input weights are applied (input layer in fig.1). Exploiting phase modulated masks, the dynamical behavior of the two lasers is enhanced at high frequencies, thus reducing the value of $\theta$ to the limits of the current ADCs capabilities [9]. Also, adopting phase modulated masks, the dependence of the processing speed from the relaxation oscillation frequency (photon lifetime depended) of the laser can be suppressed. The system of two mutually coupled lasers comprises the reservoir, which has the role of expanding the signal in a higher dimensional space through its rich non-linear dynamics (reservoir layer in fig. 1). Mutual coupling strength is controlled with the use of a VOA. Here, a gain/absorption element could be considered, in order to properly adjust feedback parameter also compensating for in-path losses attributed to the waveguide and phase modulator [18]. The output of each laser is directed to a PD and then to an ADC. The digitized outputs of the two lasers are weighted with the use of linear regression training algorithms in a similar manner as in other reservoir computing works based on delayed systems [6], [7] so as to extract the output signal (output layer in fig. 1). The ADCs sampling rate defines the minimum temporal distance $\theta$ between the virtually distributed nodes, with a reasonable value for $\theta$ being equal to 20 ps and approaching at minimum 10 ps when $T$ converges to symbol rate considering state of the art sampling systems. Before injected in the system, data must be sampled and held for a period of $T$ such that each injected symbol is spread across the delay between the two lasers. The ratio of the input signal's symbol rate and the period $T$ results in a stretching factor that defines the speed penalty of the system. For 25 Gbaud signals, there would be no speed penalty for $T = T_{sym} = 40$ ps, where $T_{sym}$ is the symbol period.

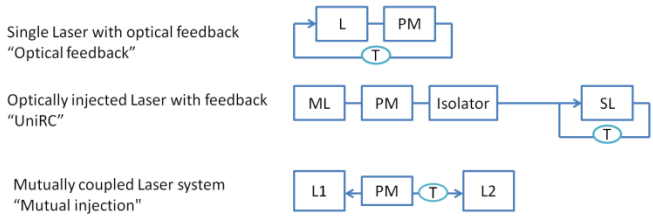

Fig. 2. Block diagram of the three different schemes implementing Time-Delayed RC. From top to bottom we see the optical feedback, the uni-directional injection locking combined with optical feedback at the slave and the mutual injection locking architectures. PM stands for Phase Modulator, L for Laser, ML for Master Laser and SL for Slave Laser.

Generally, time delayed RC with semiconductor lasers as non-linear nodes can be realized in three ways (see fig. 2). Firstly, with a single laser subjected to optical feedback, via current modulation or modulation inside the feedback cavity [17]. This scheme is the simplest, but also is less efficient performance-wise due to the lower bandwidth it can support in principle which translates to useful $\theta$ values of longer duration. On the other hand, it is compact as it requires only one laser easily fabricated in a chip. A second variant is the scheme based on the unidirectionally injected laser, with larger bandwidth and performance capabilities but also with more demanding setup and integration difficulties related to the immaturity of integrated optical isolators. The third scheme of mutually coupled lasers [10], [15], proposed here for the first time with phase modulation and without external injection, tries to combine the merits of the aforementioned schemes. In this work, we call for the sake of simplicity, the first one "optical feedback", the second "uniRC" and the third "mutual injection".

We model the three systems with the rate equations for the normalized complex slowly varying envelope of the optical field $E$ and carrier number $N$ [12]:

$$\frac{dE_{1,2}}{dt} = \frac{1+i\alpha_{1,2}}{2}\left[G_{s1,2} - \frac{1}{t_{ph1,2}}\right]E_{1,2}$$
$$+ \frac{kE_{2,1}(t-T)e^{-i[\pm 2\pi\Delta f t + \omega_{2,1}T - \varphi_m(t-T)]}}{t_{rt1,2}}$$
$$+ \sqrt{2\beta N_{1,2}}\xi \quad (1)$$

$$\frac{dE_{uni}}{dt} = \frac{1+i\alpha_{uni}}{2}\left[G_{s,uni} - \frac{1}{t_{ph,uni}}\right]E_{uni}$$
$$+ \frac{k_{inj}}{t_{rt}}E_{ext}(t)e^{-i[2\pi\Delta f t - \varphi_m(t)]}$$
$$+ \frac{k_f E_{uni}(t-T)e^{-i\omega_0 T}}{t_{rt}} + \sqrt{2\beta N_{uni}}\xi \quad (2)$$

$$\frac{dE_f}{dt} = \frac{1+i\alpha_f}{2}\left[G_{s,f} - \frac{1}{t_{ph,f}}\right]E_f$$
$$+ \frac{k_f E_f(t-T)e^{-i[\omega_0 T - \varphi_m(t-T)]}}{t_{rt}}$$
$$+ \sqrt{2\beta N_f}\xi \quad (3)$$

TABLE I
NUMERICAL MODEL PARAMETERS

| Symbol | Parameter | Value |
|---|---|---|
| $g$ | Differential gain parameter | $1.2\times10^{-8}$ ps$^{-1}$ |
| $s$ | Gain saturation coefficient | $5\times10^{-7}$ |
| $\beta$ | Spontaneous emission rate | $1.5\times10^{-10}$ ps$^{-1}$ |
| $t_n$ | Carrier lifetime | 2 ns |
| $N_0$ | Transparency Carrier Number | $1.5\times10^8$ |
| $a$ | Linewidth enhancement factor | 2 |
| $T$ | External cavity round-trip time | From 40 ps to 960 ps |
| $t_{ph}$ | Photon line-time | 2ps |
| $t_{rt}$ | Cavity round-trip time | 8ps |
| $\omega_0$ | Central oscillation frequency | $1.206\times10^{15}$ rad/sec |
| $\theta$ | Node separation | 10 ps/ 20 ps |
| $k$ | Injection parameter | From 0.03 to 0.55 |
| $I$ | Bias current | From 20mA to 70mA |

$$\frac{dN_i}{dt} = \frac{I_i}{q} - \frac{N_i}{t_{n,i}} - G_i |E_i|^2 \quad (4)$$

$$G_i = \frac{g[N_i - N_0]}{1 + s|E_i|^2} \quad (5)$$

Here equation (1) refers to the mutual injection scheme, eq. (2) is for the external injection with feedback and eq. (3) for the single laser with feedback. Besides the parameters appearing in table I [16], [18], $G_i$ is the gain per time unit of the $i$-th laser, $\Delta f$ is the frequency detuning between the two lasers when two lasers participate in the RC system (uniRC and mutual injection cases), $I_i$ the bias current of the $i$-th laser and $\xi(t)$ the spontaneous emission noise (Langevin forces) modelled as a complex Gaussian process [19]. $E_i$ corresponds to the field of the $i$-th laser, modulated at phase with the term $\varphi_m(t) = m(t)d(t)$, where $m(t)$ is the mask signal and $d(t)$ the distorted and noise impaired PAM-4 signal and $N_i$ is the carrier number. The amplitude of the phase modulation term $\varphi_m(t)$ takes values from 0 to $\pi$. In the simulations that follow we consider identical laser parameters for the two lasers of mutual injection scheme, having in mind that they are developed on the same wafer. Nevertheless, we examined the laser system with parameter mismatch between the two lasers and the optimal performance was more or less the same, thus the laser system is tolerant to parameter mismatch.

We benchmark the three schemes as dispersion compensators in direct detect transmission links. This is a very important and difficult task which attracts the interest of optical communications experts who apply techniques of digital signal processing, machine learning and lately neuromorphic computing [12], [16], [20], [21]. Especially, photonic RC in the form of hardware accelerators could substantially reduce the energy consumption attributed to DSP, which in the case of multi-km C-band transmission is fairly above 1 Watt. For comparison, a third order Volterra Non-Linear Equalizer (VLNE) with 40 2nd order taps and 10 3rd order taps may be capable to mitigate the ISI induced power fading of such links, but its complexity and consumption makes its use prohibitive. The proposed setup has 2 DFB nodes, 2 PDs of 25 GHz and two ADCs, which in the practical scenario of 50 Gsa/s sampling rate are expected to consume much less than the high-speed DSP of a complex nonlinear equalizer.

We simulate, with the integration of Nonlinear Schrodinger equation (NLSE) using the Split-step Fourier method, the transmission of 25 Gaud PAM-4 signals in a range of 40 km to 100 km. Signal propagation in our model is governed by Manakov equations [22]. Then, we improve through the RC system the bit-error rate (BER) performance of the severely distorted signal as a result of chromatic dispersion and its interplay with the nonlinearity of the square law detector. In this process, we consider all types of noises (thermal noise [23], Langevin forces in rate-equation modelling [19], shot noise [23]) as gaussian distributions of relevant standard deviation and ADCs with 8-bit resolution. Also, a gain element with 5 dB noise figure is taken into account as an optical pre-amplifier at long distances (> 50 km) [23]. In simulations, a typical launch power of 0 dBm is assumed, with the resulting electrical signal to noise ratio (SNR) to be approximately 27 dB for 50 km of transmission when no optical amplifier is considered. From the output of the virtual nodes we estimate the initially transmitted signal through Tikhonov regularization. The training is made upon a batch of symbols. The input $x_t$ is the distorted symbol sequence which has the following form $x_{t,m} = [x_{t-l}, \ldots, x_{t-1}, x_t, x_{t+1}, \ldots, x_{t+l}]$, where $m$ stands for the overall length of the batch which is equal to $m = 2l+1$. So, for the symbol at time $t$ we also launch $l$ preceding and $l$ succeeding symbols in order to track intersymbol dependencies induced by the dispersive channel. The length of this batch ranges between 11 and 51 symbols reflecting channel memory for the transmission distances considered in this paper. Using 20000 symbols as the training set and 100000 for testing, we achieve stable BER performance and reliable detection of BER values in the order of $10^{-3}$ which is close to the desired BER value of conventional Forward Error Correction (FEC) limit of $3.8 \times 10^{-3}$.

## III. RESULTS

### A. Mutual Injection Scheme

Before performing simulations, it is critical to have a picture of the dimensions of the proposed chip. Chip dimensions, mainly determined by the distance between the two lasers, also define the number of RC nodes and speed penalty. A distance of $d$ between the two lasers corresponds to a delay equal to $T = d/v_g$. In III-V or IV materials, for $d = 1$ cm, $T \sim 120$ ps. For $\theta = 20$ ps, then a 1 cm long chip can provide 6 virtual nodes per laser output, thus 12 nodes in total. For these dimensions the speed penalty factor equals to $T/T_{sym} = 3$. As we increase $d$, we increase the number of nodes sacrificing speed. Based on the results of [12] we consider that $T = 240$ps corresponds to a trade-off between computational performance and compactness/speed penalty. $T = 240$ ps or 24 node setup translates to a 2 cm long chip and a speed penalty factor of 6. At this point, it must be stressed out that the optimization of photonic RC systems is a time-consuming process, especially when more than one physical nodes are involved. Usually, an extensive mapping of the performance as a function of the main hyperparameters is performed. In this work we follow this methodology, by performing extensive mapping of the performance with respect to the two basic parameters, that is laser frequency detuning and injection strength, while the third hyperparameter (i.e the bias currents) is studied in two principal scenarios: equal and unequal biasing. Equal biasing is a promising technique to offer broadband chaos generation [24], however such a fast route to complex chaos is not preferred in RC applications. On the other hand, unequal biasing can in principle provide a method to better control non-linear dynamics and differentiate the outputs of the two lasers. Laser frequency can be easily tuned either adjusting the bias current or the temperature, provided that either the two lasers are on different chips or they exhibit different coefficient of wavelength shift as a function of temperature variations (nm/$^0$C). In our case, since the two lasers are hosted on the same chip, the safest way to detune them is to properly adjust their bias currents. A small change of bias current could easily detune the laser a few GHz without changing substantially its dynamical properties. Moreover, the injection strength can be easily controlled by varying the voltage of the VOA/gain section.

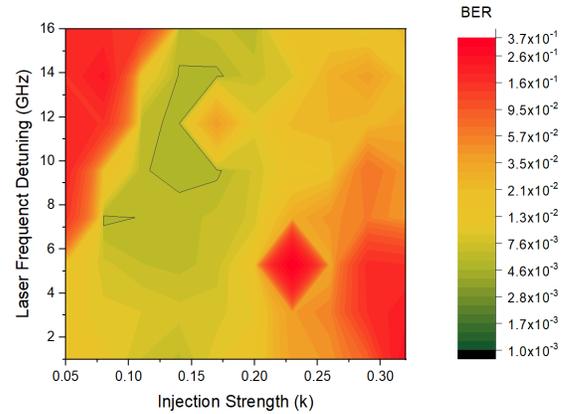

Fig. 3. BER results versus laser frequency detuning $\Delta f$ and injection parameter $k$ for 50 km of transmission distance. $T = 240$ ps. The two Lasers are equally biased at 60 mA. Identical results were taken for negative $\Delta f$.

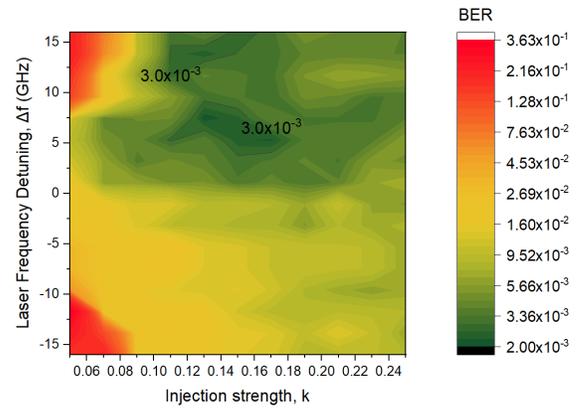

Fig. 4. BER results versus laser frequency detuning $\Delta f$ and injection parameter $k$ for 50 km of transmission distance. The two Lasers are unequally biased at 70 mA and 20 mA respectively. BER values slightly below FEC limit of $3 \times 10^{-3}$ can be obtained.

We start the numerical evaluation by considering that both lasers have the same bias current. We map performance by calculating BER of the compensated by RC signal after 50 km of transmission as a function of laser detuning and injection parameter $k$. Without RC and with the use of a linear equalizer for instance a feed forward equalizer, BER is worse than 0.1 as it will be shown extensively later. Both lasers are biased at 60mA, a current almost four times their threshold. We have chosen a high current value, as higher currents support higher frequency dynamics which are indispensable in this specific high-speed transmission application where $\theta$ must be reduced as a much as possible [9]. Moreover, high current values increase signal to noise ratio [7]. Fig. 3 shows how BER evolves in different operating regimes. We observe that with equally biased and identical, in terms of intrinsic parameters, lasers the best BER results occur when laser detuning $\Delta f$ resides from 6 to 16 GHz and $k$ between 0.1 and 0.18 respectively. Within the range where optimal performance is observed, BER values are around $6 \times 10^{-3}$ approaching FEC limit whilst the correlation of the two laser outputs approaches 90%. Outside this region, either the nonlinear dynamics are not appropriate for RC and/or the correlation of the two outputs approaches 99%. For instance, for higher $k$ values, the laser outputs exhibit rich chaotic nonlinear dynamics [24], which are not appropriate for RC applications.

In order to further decorrelate the dynamics of the two lasers, we bias them at different currents. In this second scenario, $I_1 = 70$ mA and $I_2 = 20$ mA. The results are depicted in fig. 4. It becomes obvious that two mutually injected lasers with different bias conditions constitute a RC system with better compensation capabilities (BER $< 3 \times 10^{-3}$) in an extended range for $k$ (0.08 to 0.22) and $\Delta f$ (2 to 16 GHz) values when compared to the previous case of identical biasing. The main reason for this moderate improvement is that the two laser outputs become more decorrelated than before. In the region of optimal performance, we calculate a correlation coefficient in the order of 0.6, which substantially increases the number of useful nodes for computing compared to the almost 90% correlated nodes of the two lasers under equal biasing. The decorrelation increases as the currents of the two lasers diverge. Moreover, unequal biasing provides a better control of the dynamical behavior of the overall system which is reflected to the wide range of operating conditions that provide low BER performance (close to FEC limit) in contrast to equal biasing. Another difference is that for unequal biasing, the sign of $\Delta f$ affects RC performance with positive $\Delta f$ being the favorite operating regime as laser2 response to laser1 injection does not exhibit a symmetrical bandwidth with respect to laser2 central frequency due to $a$ parameter. It must be pointed out that the system is phase sensitive due to the small cavity lengths especially when the two lasers are equally biased. In all cases, we have determined the required phase shift that is needed in order to achieve the best possible BER.

### B. Comparison With Similar Schemes

For the mutual injection scheme, we choose as optimal regime the one corresponding to $I_1 = 70$ mA, $I_2 = 20$ mA, $k = 0.14$ and $\Delta f = 9$ GHz as dictated by fig. 4 results. The RC scheme based on optical feedback and external optical injection of [16], has been optimized, using eq. (4), with injection strength $k = 0.85$ and feedback strength $k_f = 0.08$, $\Delta f = 0$ GHz, and the master and slave laser biased two times the threshold ($\sim$30 mA). Although in [16], the authors operate the slave laser close to the threshold, here, with the use of phase modulation we identified similar optimal performance even when the slave laser is fairly above threshold. Moreover, phase modulation alleviates the need for frequency detuning between master and slave as phase modulation is equivalent to frequency modulation. For the single laser with feedback, we draw the contour plot of fig. 5 in order to determine the optimal conditions of feedback strength and bias current with respect to BER performance which also shows that phase modulation permits best results to occur at high currents where higher modulation bandwidth and smaller $\theta$ values can be achieved.

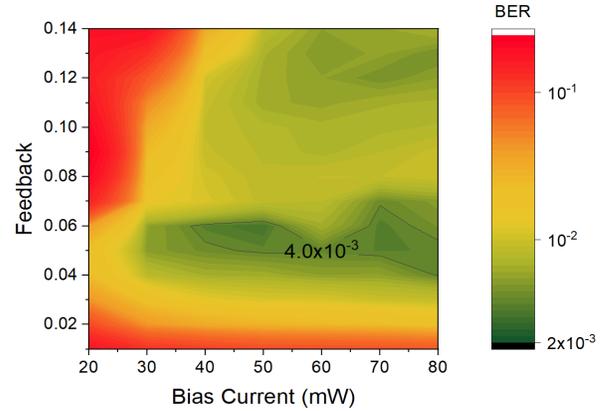

Fig. 5. BER results versus Bias current and Feedback strength for the single laser with phase modulation inside cavity (Optical Feedback). With a delay cavity of 480 ps and 45 km transmission this scheme can provide adequate results for feedback strength around 0.05 and bias current above 30 mW. We choose this best regime for the comparison with the other schemes.

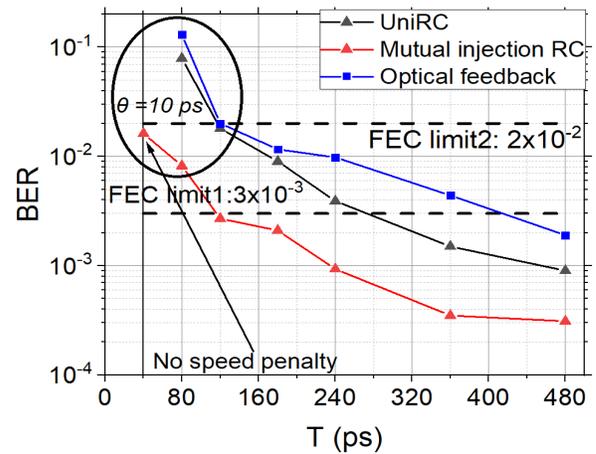

Fig. 6. Comparison between uniRCs and mutual injection RC architectures for 45 km of transmission in single mode fiber as a function of $T$. $\theta = 20$ ps in all cases, except from those indicated in the circle, where $\theta = 10$ ps.

The first comparison focuses on how RC behaves as a function of $T$ values when the PAM-4 signals have undergone 45 km of transmission. For the studies of fig. 6, 7 we also introduce open FEC limit ($2 \times 10^{-2}$), which has higher tolerance to noise but at expense of an increased overhead of approximately 15% compared to the 7% overhead of the conventional FEC. Delay loop is the limiting factor in terms of compactness, and processing speed. In order to approach very small delays, close to 40 ps, which reassure real-time processing with no speed penalty, we have lowered $\theta$ to 10 ps. Such a small $\theta$ value is meaningful for strong optical injection, phase modulation and high bias current that all together reassure high bandwidth operation and thus small temporal separation of the nodes. As fig. 6 shows, in the case of equal delay loops, the systems relying on optical injection (mutual injection and uniRC) perform better than the system relying on optical feedback as expected. Between mutual injection and uniRC, the mutual injection system proposed in this work performs better, due to the fact it provides almost double the number of nodes at the same T value. One can easily observe that a uniRC system becomes almost equivalent to mutual injection when its delay has the double value compared to that in the mutual injection case, thus the difference in their performance is a matter of the number of virtual nodes. One could increase the number of nodes by increasing the number of slave lasers [6], however the challenges relating to optical isolation in monolithic integrated implementations of UniRC still remain. Thanks to optical injection and the double number of nodes offered using two physical nonlinear nodes, it is important to note that BER values close to open FEC limit are obtained even for $T = 40$ ps, which means that the proposed RC can be used for real time processing of 25 Gbaud signals offering better performance than linear feed forward equalizers (FFE) as depicted in fig. 7. BER values below $3 \times 10^{-3}$ can be obtained for $T = 120$ ps, that is for processing rate approaching 10 Gsa/sec which is quite important for a wide range of other applications except optical communications.

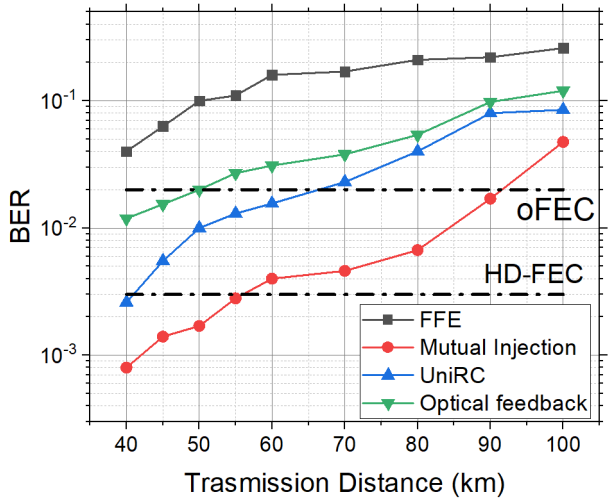

Fig. 7. Comparison between a linear equalizer, uniRCs and mutual injection RC for different transmission distances of the PAM-4 signal. Mutual injection scheme outperforms the other techniques.

We then compare the three systems for $T = 240$ ps (speed penalty = 6) in terms of their computational efficiency in mitigating impairments over different distances. In this case,
mutual injection setup achieves the best results due to the double number of nodes it provides. On the other hand, optical feedback scheme is slightly more compact, as it can provide the same optical delay with mutual injection using half the size of the chip. That is, with 1 cm long chips optical feedback can achieve $T = 240$ ps whilst mutual injection needs 2 cm long chips for the same delay. Finally, UniRC also relying on optical feedback has the advantage of a shorter chip length for the same delay compared to mutual injection, however it is still technologically immature in terms of monolithic integration as already explained. Mutual injection system provides BER $< 3 \times 10^{-3}$ (HD-FEC limit) for distances shorter than 55 km. UniRC accomplishes the same performance for distances shorter than 45 km. As far as oFEC is concerned, mutual injection is successful for up to 90 km of transmission, whilst single laser schemes can hardly reach 60 km. Thus, the proposed RC system can substantially enhance the reach of PAM-4 directly detected signals. The linear equalizer (FFE) is by far inferior compared to the two RC schemes as already shown in many papers in the literature [12], [16], showing once more that RC has the capacity to restore directly detected dispersion impaired signals at very high processing rates. The fact that mutual injection scheme can provide adequate results even with 80 or 40 ps delay loops opens a new path for delay-based systems, with operation beyond 20 Gsa/s being feasible. Finally, it is evident that mutual injection setup provides the best tradeoff between compactness and performance.

## IV. CONCLUSION

In conclusion, we proposed a novel RC system based on mutually coupled lasers that can be monolithic integrated together with high-speed electronics, thus constituting an opto-electronic RC system of high computational efficiency at speeds beyond 10 Gsa/s. The system exploits phase modulation that can provide better response at high frequencies, electrical injection and doubling of RC nodes under proper operating conditions which decorrelate the outputs of the two mutually coupled lasers. The scheme performs better for unequally biased lasers and frequency detuning values in the range of 2-16 GHz. In this regime, simulations show that the system enhances the transmission reach of PAM-4 25 Gbaud signals in intensity modulation direct detection optical communication systems showing that even 90 km of transmission can be accomplished if Open FEC is employed for error correction. The mutual injection scheme outperforms the configurations based on the externally injected laser with feedback and a single laser under optical feedback when compared at the same RC characteristic delay.

## REFERENCES

[1] J. Zhu, T. Zhang, Y. Yang, and R. Huang, "A comprehensive review on emerging artificial neuromorphic devices," *Appl. Phys. Rev.*, vol. 7, no. 1, Mar. 2020, Art. no. 011312.
[2] B. J. Shastri *et al.*, "Photonics for artificial intelligence and neuromorphic computing," *Nature Photon.*, vol. 15, no. 2, pp. 102–114, Feb. 2021.
[3] H. Jaeger, "The 'echo state' approach to analysing and training recurrent neural networks-with an erratum note," German Nat. Res. Center Inf. Technol. GMD, Bonn, Germany, Tech. Rep. 148, 2001.
[4] K. Vandoorne *et al.*, "Toward optical signal processing using photonic reservoir computing," *Opt. Exp.*, vol. 16, no. 15, pp. 11182–11192, 2008.

[5] L. Appeltant et al., "Information processing using a single dynamical node as complex system," *Nature Commun.*, vol. 2, no. 1, Sep. 2011, Art. no. 468.

[6] C. Sugano, K. Kanno, and A. Uchida, "Reservoir computing using multiple lasers with feedback on a photonic integrated circuit," *IEEE J. Sel. Top. Quantum Electron.*, vol. 26, no. 1, Jan./Feb. 2020, Art. no. 1500409.

[7] K. Harkhoe, G. Verschaffelt, A. Katumba, P. Bienstman, and G. Van der Sande, "Demonstrating delay-based reservoir computing using a compact photonic integrated chip," *Opt. Exp.*, vol. 28, no. 3, p. 3086, 2020.

[8] J. Vatin, D. Rontani, and M. Sciamanna, "Enhanced performance of a reservoir computer using polarization dynamics in VCSELs," *Opt. Lett.*, vol. 43, no. 18, pp. 4497–4500, 2018.

[9] R. M. Nguimdo, G. Verschaffelt, J. Danckaert, and G. Van der Sande, "Fast photonic information processing using semiconductor lasers with delayed optical feedback: Role of phase dynamics," *Opt. Exp.*, vol. 22, no. 7, pp. 8672–8686, 2014.

[10] Y.-S. Hou, G.-Q. Xia, E. Jayaprasath, D.-Z. Yue, W.-Y. Yang, and Z.-M. Wu, "Prediction and classification performance of reservoir computing system using mutually delay-coupled semiconductor lasers," *Opt. Commun.*, vol. 433, pp. 215–220, Feb. 2019.

[11] K. Harkhoe and G. Van der Sande, "Delay-based reservoir computing using multimode semiconductor lasers: Exploiting the rich carrier dynamics," *IEEE J. Sel. Top. Quantum Electron*, vol. 25, no. 6, Nov./Dec. 2019, Art. no. 1502909.

[12] A. Bogris, C. Mesaritakis, S. Deligiannidis, and P. Li, "Fabry–Pérot lasers as enablers for parallel reservoir computing," *IEEE J. Sel. Topics Quantum Electron.*, vol. 27, no. 2, pp. 1–7, Mar. 2021.

[13] D. Yue et al., "Performance optimization research of reservoir computing system based on an optical feedback semiconductor laser under electrical information injection," *Opt. Exp.*, vol. 27, no. 14, pp. 19931–19939, 2019.

[14] A. Fragkos, A. Bogris, D. Syvridis, and R. Phelan, "Amplitude noise limiting amplifier for phase encoded signals using injection locking in semiconductor lasers," *J. Lightw. Technol.*, vol. 30, no. 5, pp. 764–771, Mar. 1, 2012.

[15] S. Ortín and L. Pesquera, "Reservoir computing with an ensemble of time-delay reservoirs," *Cogn. Comput.*, vol. 9, no. 3, pp. 327–336, 2017.

[16] A. Argyris, J. Bueno, and I. Fischer, "Photonic machine learning implementation for signal recovery in optical communications," *Sci. Rep.*, vol. 8, no. 1, Dec. 2018, Art. no. 8487.

[17] R. M. Nguimdo, E. Lacot, O. Jacquin, O. Hugon, G. Van der Sande, and H. G. de Chatellus, "Prediction performance of reservoir computing systems based on a diode-pumped erbium-doped microchip laser subject to optical feedback," *Opt. Lett.*, vol. 42, no. 3, p. 375, 2017.

[18] A. Argyris, M. Hamacher, K. E. Chlouverakis, A. Bogris, and D. Syvridis, "Photonic integrated device for chaos applications in communications," *Phys. Rev. Lett.*, vol. 100, no. 19, May 2008, Art. no. 194101.

[19] C. Henry, "Theory of spontaneous emission noise in open resonators and its application to lasers and optical amplifiers," *J. Lightw. Technol.*, vol. 4, no. 3, pp. 288–297, Mar. 1986.

[20] C. Huang et al., "Demonstration of photonic neural network for fiber nonlinearity compensation in long-haul transmission systems," in *Proc. OFC*, San Diego, CA, USA, 2020, pp. 1–3.

[21] S. Zhang et al., "Field and lab experimental demonstration of nonlinear impairment compensation using neural networks," *Nature Commun.*, vol. 10, no. 1, p. 3033, Dec. 2019.

[22] D. Marcuse, C. R. Manyuk, and P. K. A. Wai, "Application of the Manakov-PMD equation to studies of signal propagation in optical fibers with randomly varying birefringence," *J. Lightw. Technol.*, vol. 15, no. 9, pp. 1735–1746, Sep. 1997.

[23] R. Ramaswami, K. Sivarajan, and G. Sasaki. *Optical Networks: A Practical Perspective*. San Mateo, CA, USA: Morgan Kaufmann, 2009.

[24] L. Qiao et al., "Generation of flat wideband chaos based on mutual injection of semiconductor lasers," *Opt. Lett.*, vol. 44, no. 22, pp. 5394–5397, 2019.

**Kostas Sozos** received the B.S. degree in physics from the University of Patras in 2018 and the M.Sc. degree in microsystems and nanodevices from the National Technical University of Athens in 2020. He is currently pursuing the Ph.D. degree in neuromorphic photonics with the University of West Attica, under the supervision of Prof. Adonis Bogris. His main research interests include photonic neuromorphic computing, pattern recognition, and optical communications.

**Charis Mesaritakis** received the B.S. degree in informatics from the Department of Informatics and Telecommunications, National and Kapodistrian University of Athens, in 2004, the M.Sc. degree in microelectronics from the Department of Informatics and Telecommunications, National and Kapodistrian University of Athens, and the Ph.D. degree in quantum dot devices and systems for next generation optical networks from the Photonics Technology and Optical Communication Laboratory, National and Kapodistrian University of Athens, in 2011. He is currently an Associate Professor with the Department of Information and Communication Systems Engineering, University of the Aegean, Greece. He has actively participated as a research engineer/technical supervisor in more than ten EU-funded research programs (FP6-FP7-H2020) targeting excellence in the field of photonic neuromorphic computing, cyber-physical security, and photonic integration. He is the author or coauthor of more than 60 papers in highly cited peer-reviewed international journals and conferences and two international book chapters, whereas he serves as a Regular Reviewer for IEEE. In 2012, he was awarded a European scholarship for post-doctoral studies (Marie Curie FP7-PEOPLE IEF) in the joint research facilities of Alcatel-Thales-Lucent in Paris–France, where he worked on intra-satellite communications.

**Adonis Bogris** was born in Athens. He received the B.S. degree in informatics, the M.Sc. degree in telecommunications, and the Ph.D. degree from the National and Kapodistrian University of Athens, Athens, in 1997, 1999, and 2005, respectively. His Ph.D. thesis was on all-optical processing by means of fiber-based devices. He is currently a Professor with the Department of Informatics and Computer Engineering, University of West Attica, Greece. He has authored or coauthored more than 150 articles published in international scientific journals and conference proceedings, and participated in plethora of EU and national research projects. His current research interests include high-speed all-optical transmission systems and networks, nonlinear effects in optical fibers, all-optical signal processing, neuromorphic photonics, mid-infrared photonics, and cryptography at the physical layer. He serves as a Reviewer for the IEEE journals.